\documentclass[aps,prb,twocolumn,groupedaddress]{revtex4}
\usepackage{graphicx,bm,amssymb}

\begin{document}

\title{Spin relaxation and anticrossing in quantum dots: Rashba versus Dresselhaus spin-orbit coupling}

\author{Denis V. Bulaev and Daniel Loss}
\affiliation{Department of Physics and Astronomy, University of Basel, Klingelbergstrasse 82, CH-4056 Basel, Switzerland}

\date{\today}

\begin{abstract}
The spin-orbit splitting of the electron levels in a two-dimensional quantum dot in a perpendicular magnetic field is studied. It is shown that at the point of an accidental degeneracy of the two lowest levels above the ground state the Rashba spin-orbit coupling leads to a level anticrossing and to mixing of spin-up and spin-down states, whereas there is no mixing of these levels due to the Dresselhaus term. We calculate the relaxation and decoherence times of the three lowest levels due to phonons. We find that the spin relaxation rate as a function of a magnetic field exhibits a cusplike structure for Rashba but not for Dresselhaus spin-orbit interaction.
\end{abstract}


\maketitle

\section{Introduction}
Recent years have seen an increasing interest in the spin properties of nanostructures \cite{ALS}. Manipulation and readout of spins in solids could open the way to the development of a generation of electronic devices such as spin transistors, spin filters, and spin memory devices. In addition, the spin of an electron confined to a quantum dot (QD) is a promising candidate for a quantum bit \cite{LV}. Owing to the zero dimensionality of QDs, the electronic orbital states are quantized and the electron spin states are very stable due to a substantial suppression of spin-flip mechanisms \cite{PBS,KN}. Progress in nanotechnology has allowed the fabrication of QDs with desirable electronic and spin properties \cite{F2001,FTH2001,F2002,H2003,Hanson2004}. However, only recently it has been possible to measure the spin of an electron in a QD. A single electron spin has been detected by magnetic resonance force microscopy \cite{RBMCh} and  the readout of an individual electron spin in a QD via pulsed relaxation measurements \cite{E2004} and optical orientation experiments \cite{K2004} have been reported. In these experiments, an external magnetic field was used to distinguish spin-up and spin-down states split by the Zeeman energy. Spin relaxation measurements between Zeeman levels in a QD \cite{E2004,K2004} confirm the theoretical predictions that spin-flip relaxation in a QD is suppressed with respect to a bulk structure \cite{PBS,KN,GKL}. Indeed, very long single-spin relaxation times have been observed: up to $0.85\:$ms in two-dimensional (2D) GaAs QDs \cite{E2004}, and up to $20\:$ms in self-assembled GaInAs QDs \cite{K2004}. The spin relaxation is expected to be dominated by hyperfine interactions with the nuclei at magnetic fields below $0.5\:$T \cite{EN,KLG,CL} and  by spin-orbit (SO) interactions for magnetic fields of about $1\:$Tesla (see Ref.~\onlinecite{KN}) and for higher magnetic fields (see Ref.~\onlinecite{GKL}). In general, the SO interaction consists of two distinct contributions:
the Dresselhaus SO coupling  \cite{D} which is due to bulk inversion asymmetry of the lattice and the Rashba SO coupling  \cite{R} which is due to structure inversion asymmetry along the growth direction. Both of these SO terms result in the splitting of electron energy levels and in the mixing of the electron spin states. The latter makes  spin-flip relaxation between Zeeman levels possible, for example, due to the phonon scattering. Note that usually it is not simple to separate these two SO mechanisms and estimate the relative contributions of each SO term. In experiments, to obtain information about one of the SO couplings, normally the other is neglected \cite{LMFS,C2002,Zumbuehl}.
 This leads to a lack of precision in estimates of the SO coupling strength and to a neglect of the effects of the interplay of the Rashba and the Dresselhaus SO couplings \cite{GKL,G2004,AF}. Hence, it is very important to find a way to separate these SO mechanisms, to increase our understanding of the SO relaxation processes, and to improve predictions of the spin properties of nanostructures. 
It is well known \cite{Silva}  that for 2D quantum wells the different SO couplings can be distinguished
experimentally \cite{Jusserand,G2004,Miller}
via detection of the
associated anisotropy of the spin splitting in the conduction band.
In contrast, such a detection is not possible in QDs since the spin splitting of the levels, 
being quadratically in the SO coupling,
is isotropic. Still, as we point out now, the SO couplings in QDs can be distinguished via their associated 
spin relaxation rates since they strongly differ due to different level mixing properties.

In this paper, the electron energy spectrum and the spin relaxation for a 2D QD in magnetic fields perpendicular to the QD surface are studied.  Level anticrossing \cite{DUM} (due to the SO coupling), at a point of accidental level degeneracy (due to the interplay between the orbital and magnetic confinement), is analytically investigated. 
This anticrossing is caused by the Rashba SO term only, leading to a cusp structure in the magnetic-field dependence of the spin relaxation rate, whereas the spin relaxation rate due to the Dresselhaus SO coupling is a monotonic function of magnetic field in this region. This qualitative difference in the spin relaxation for different SO couplings can serve to extract the different contributions in SO coupling.

\section{Model and energy spectrum}
We consider a 2D isotropic QD with parabolic lateral confinement potential. An external magnetic field is applied perpendicularly to the surface of the QD. The Hamiltonian of this system reads
\begin{equation}
\label{eq:H0}
H_0=\frac{\mathbf{P}^2}{2m^*}+\frac12m^*\omega_0^2\left(x^2+y^2\right)+\frac12g\mu_BB\sigma_z,
\end{equation}
where $\mathbf{P}=\mathbf{p}+(|e|/c)\mathbf{A(r)}$, $\mathbf{A(r)}=(B/2)(-y,x,0)$ is the vector potential in the symmetric gauge, $\omega_0$ is the characteristic confinement frequency, and $\bm{\sigma}=(\sigma_x,\sigma_y,\sigma_z)$ is the vector of the Pauli matrices.

The SO interaction is taken into account by adding the linear Dresselhaus \cite{D,footnote}  and Rashba \cite{R} terms for conduction band electrons in a [001] two-dimensional electron gas (2DEG),
\begin{equation}
\label{eq:HDR}
H_D=\beta(-\sigma_xP_x+\sigma_yP_y),\ H_R=\alpha(\sigma_xP_y-\sigma_yP_x).
\end{equation}
The axes $x$, $y$, and $z$ are aligned along the principal crystallographic axes of GaAs.

It is convenient to introduce new phase coordinates ($q_1,\ q_2,\ p_1,\ p_2$) which connected to the previous ones ($x,\ y,\ p_x,\ p_y$) by the following formula \cite{GMSh}:
\begin{eqnarray*}
x&=&\frac{1}{\sqrt{2\Omega}}\left(\sqrt{\omega_1}q_1+\sqrt{\omega_2}q_2\right),\\
y&=&\frac{1}{m^*\sqrt{2\Omega}}\left(\frac{p_1}{\sqrt{\omega_1}}-\frac{p_2}{\sqrt{\omega_2}}\right),\\
p_{x}&=&\sqrt{\frac{\Omega}{2}}\left(\frac{p_1}{\sqrt{\omega_1}}+\frac{p_2}{\sqrt{\omega_2}}\right),\\
p_{y}&=&m^*\sqrt{\frac{\Omega}{2}}\left(-\sqrt{\omega_1}q_1+\sqrt{\omega_2}q_2\right),
\end{eqnarray*}
where 
\[
\Omega=\sqrt{\omega_0^2+\omega_c^2/4},\ \omega_{1,2}=\Omega\mp\frac{\omega_c}{2}. 
\]
Here $\omega_c=|e|B/m^*c$ is the cyclotron frequency.
In the new phase coordinates, $H_0$ has the canonical form 
\begin{eqnarray}
\label{eq:Canonical}
H_0&=&\frac{p_1^2+p_2^2}{2m^*}+ \frac{m^*}{2}(\omega_1^2q_1^2+\omega_2^2q_2^2)+\frac12g\mu_BB\sigma_z.
\end{eqnarray}
In this case, $H_0$ can be considered as the Hamiltonian of two independent harmonic oscillators with hybrid frequencies $\omega_{1,2}$. Therefore, the energy spectrum and eigenstates of electrons in a QD without the SO coupling are given by
\begin{eqnarray*}
E_{nms_z}^{(0)}&=&\hbar\omega_1(n+1/2)+\hbar\omega_2(m+1/2)-\hbar\omega_Z s_z,\\
\left\langle q_1q_2|nms_z\right\rangle&=&\Phi_n(q_1\sqrt{m^*\omega_1/\hbar})\Phi_m(q_2\sqrt{m^*\omega_2/\hbar})\left|s_z\right\rangle,
\end{eqnarray*}
where $n,m=0,1,2,\ldots$, $s_z=\pm1/2$ is the electron-spin projection on the $z$-axis, $\omega_Z=|g|\mu_BB/\hbar$ is the Zeeman frequency, and  $\Phi_n(q)$ are oscillator functions. 

Let us consider the three lowest levels:
\begin{eqnarray*}
E_{00\uparrow}^{(0)}&=&\hbar\Omega-\hbar\omega_Z/2,\ E_{00\downarrow}^{(0)}=\hbar\Omega+\hbar\omega_Z/2,\\
E_{10\uparrow}^{(0)}&=&\hbar\Omega+\hbar\omega_1-\hbar\omega_Z/2.
\end{eqnarray*}
 The first level is the ground state. In the case of weak magnetic confinement ($\omega_0\gg\omega_c$), the second level is lower than the third one ($E_{00\downarrow}^{(0)}<E_{10\uparrow}^{(0)}$). However, at high magnetic fields, when the magnetic confinement is much stronger than the lateral confinement ($\omega_0\ll\omega_c$), $E_{00\downarrow}^{(0)}>E_{10\uparrow}^{(0)}$, because $E_{10\uparrow}^{(0)}\approx E_{00\uparrow}^{(0)}+\hbar\omega_0^2/\omega_c$. The condition for a crossing of the levels $E_{00\downarrow}^{(0)}$ and $E_{10\uparrow}^{(0)}$ is given by $\omega_1=\omega_Z$. 
In other words, this level crossing takes place when the magnetic length $l_B=\sqrt{\hbar/m^*\omega_c}$ is equal to $l_0[g^*(g^*+1)]^{1/4}$, where $l_0=\sqrt{\hbar/m^*\omega_0}$ is the characteristic lateral size of a QD and $g^*=|g|m^*/2m_0$. Note that the level crossing occurs at accessible magnetic fields for QDs with lateral size $l_0>15\:$nm.

Now we take SO coupling into  account and find the energy spectrum and eigenstates of electrons in a QD. For a GaAs QD the SO lengths are $\lambda_D=\hbar/m^*\beta,\ \lambda_R=\hbar/m^*\alpha\approx8\:\mu$m \cite{Zumbuehl} and are much larger than the hybrid orbital length $l=\sqrt{\hbar/m^*\Omega}$ of a QD ($\lambda_D,\ \lambda_R\gg l$). Therefore, the SO terms can be considered as small perturbations.

First we consider the Dresselhaus SO coupling [see Eq.~(\ref{eq:HDR})]. It is important to note that in first-order perturbation theory there is no SO interaction between the levels $E_{00\downarrow}^{(0)}$ and $E_{10\uparrow}^{(0)}$ due to the Dresselhaus term ($\langle00\downarrow|H_D|10\uparrow\rangle=0$). Hence we can apply standard perturbation theory for nondegenerate levels. Thus, in first-order perturbation theory, we get $E_n=E_n^{(0)}$,
\begin{eqnarray}
|1\rangle&=&  |00\uparrow\rangle +\frac{(l/\lambda_D)\omega_1}{\omega_1+\omega_Z}|10\downarrow\rangle,\\
\label{eq:State2D}
|2\rangle&=& |00\downarrow\rangle - \frac{(l/\lambda_D)\omega_2}{\omega_2-\omega_Z}|01\uparrow\rangle,\\
\label{eq:State3D}
|3\rangle&=& |10\uparrow\rangle +\frac{\sqrt{2}(l/\lambda_D)\omega_1}{\omega_1+\omega_Z}|20\downarrow\rangle.
\end{eqnarray}

\begin{figure*}
\includegraphics[clip=true,width=17.8 cm]{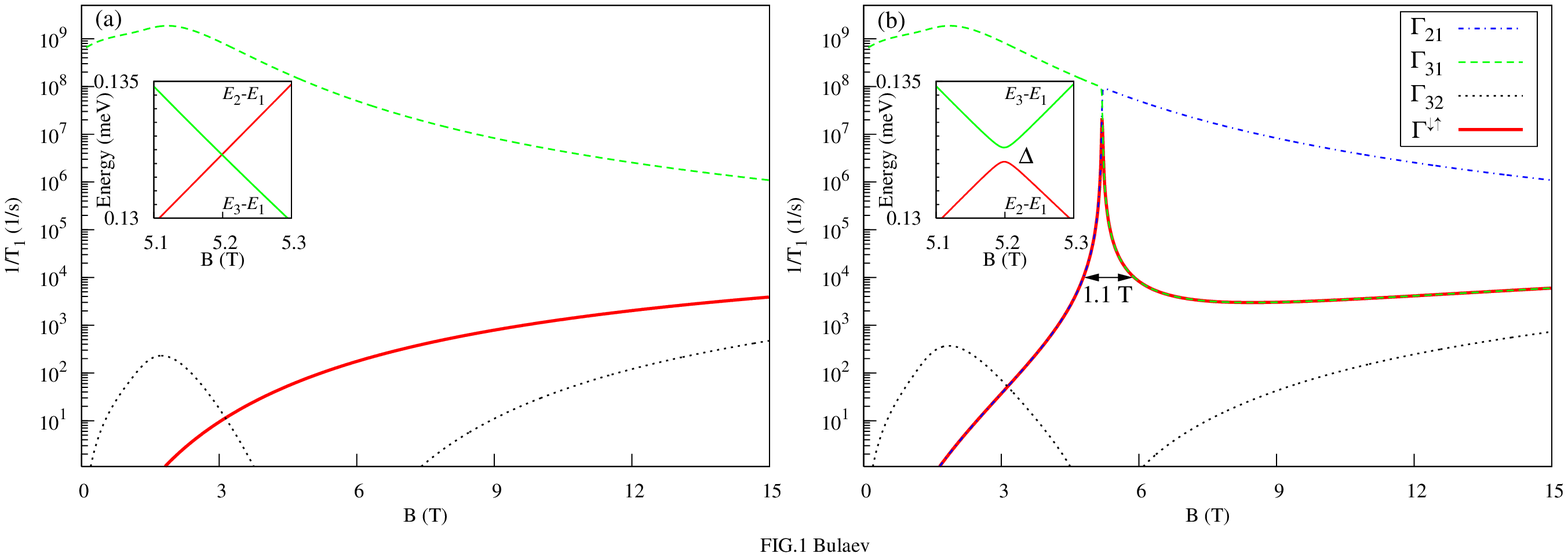}%
\caption{\label{fig:1} (color online). Contributions to the relaxation rate $1/T_1$ of phonon-induced transitions between the states $|1\rangle$, $|2\rangle$, and $|3\rangle$ of a GaAs QD with $\hbar\omega_0=1.1\:$meV and $d=5\:$nm due to (a) the Dresselhaus and (b) the Rashba SO couplings ($\lambda_D=\lambda_R=8\:\mu$m). The dashed and dot-dashed curves are orbital relaxation rates, the solid and dotted curves are the relaxation rates with a spin flip.  The crossing (a) and the anticrossing (b) of the levels $E_3$ and $E_2$ are shown in the insets. The cusplike structure of the spin relaxation curve due to the Rashba SO coupling is caused by the mixing of the spin-up and spin-down states at the anticrossing.}
\end{figure*}

Now we consider the Rashba SO coupling term. In this case, there is a SO interaction between the levels $E_{00\downarrow}^{(0)}$ and $E_{10\uparrow}^{(0)}$. Therefore, applying perturbation theory for degenerate levels, we have
\begin{eqnarray}
\label{eq:Energy1R}
E_1&=&\hbar\Omega-\frac 12\hbar\omega_Z,\ E_{2,3}=\hbar\Omega+\frac\hbar2\left(\omega_1\mp\omega_R\right),\\
|1\rangle&=&|00\uparrow\rangle -\lambda|01\downarrow\rangle,\\
\label{eq:State2R}
|2\rangle&=&\cos\frac{\gamma}{2}|00\downarrow\rangle-\sin\frac{\gamma}{2}|10\uparrow\rangle+\lambda\sin\frac{\gamma}{2}|11\downarrow\rangle,\\
\label{eq:State3R}
|3\rangle&=&\sin\frac{\gamma}{2}|00\downarrow\rangle+\cos\frac{\gamma}{2}|10\uparrow\rangle-\lambda\cos\frac{\gamma}{2}|11\downarrow\rangle,
\end{eqnarray}
where 
\begin{eqnarray}
\label{eq:wR}
\omega_R&=&\sqrt{(\omega_1-\omega_Z)^2+4(l/\lambda_R)^2\omega_1^2},\\
\nonumber
\tan\gamma&=&-2(l/\lambda_R)\omega_1/(\omega_1-\omega_Z),\\
\nonumber
\lambda&=&(l/\lambda_R)\omega_2/(\omega_2+\omega_Z).
\end{eqnarray}
As can be seen from Eq.~(\ref{eq:Energy1R}), in the case of strong lateral confinement [$\omega_1-\omega_Z\gg\left(l/\lambda_R\right)\omega_1$], $E_2=E_{00\downarrow}^{(0)}$ and $E_3=E_{10\uparrow}^{(0)}$, but in the case of strong magnetic confinement [$\omega_Z-\omega_1\gg\left(l/\lambda_R\right)\omega_1$], the levels $E_2$ and $E_3$ change places: $E_2=E_{10\uparrow}^{(0)},\ E_3=E_{00\downarrow}^{(0)}$. At the crossing point for the levels $E_{00\downarrow}^{(0)}$ and $E_{10\uparrow}^{(0)}$ ($\omega_1=\omega_Z$), $E_{2,3}=\hbar\Omega+\hbar\omega_Z/2\mp(l/\lambda_R)\hbar\omega_Z$. Therefore, \textit{the Rashba SO coupling leads to an anticrossing of the levels $E_2$ and $E_3$ at the point of accidental degeneracy of the levels $E_{00\downarrow}^{(0)}$ and $E_{10\uparrow}^{(0)}$} [see inset in Fig.~\ref{fig:1}(b)] \cite{footnote2}.
The distance between the levels $E_2$ and $E_3$ at the anticrossing is $\Delta=2(l/\lambda_R)\hbar\omega_Z$. For a GaAs QD with $\hbar\omega_0=1.1\:$meV and $\lambda_R=8\:\mu$m, this anticrossing is too small for experimental observation ($\Delta=0.5\:\mu$eV), but for an InAs  ($g\approx1$ \cite{Th} and $\lambda_R\approx0.1\:\mu$m \cite{LMFS}) QD with the same size, the anticrossing can reach $0.1\:$meV.
Note that this anticrossing features were numerically studied for narrow-gap QDs in Ref.~\onlinecite{DUM}.

Let us study the states $|2\rangle$ and $|3\rangle$. As can be seen from Eqs.~(\ref{eq:State2R}) and (\ref{eq:State3R}), if $\omega_1-\omega_Z\gg(l/\lambda_R)\omega_1$ [$\gamma=O(l/\lambda_R)$], 
\[
|2\rangle=|00\downarrow\rangle+O(l/\lambda_R),\ |3\rangle=|10\uparrow\rangle+O(l/\lambda_R). 
\]
With increasing $B$, the Zeeman energy becomes larger than $\hbar\omega_1$. In the case of $\omega_Z-\omega_1\gg(l/\lambda_R)\omega_1$,
$\gamma=\pi+O(l/\lambda_R)$ and these states change place. Therefore, the spin flips with a transition trough the anticrossing region. In the region of the anticrossing ($\gamma\approx-\pi/2$), the SO coupling of  these states due to the Rashba term becomes essential and leads to a mixing of spin-up and spin-down states: 
\begin{eqnarray*}
|2\rangle&=&({|00\downarrow\rangle} + {|10\uparrow\rangle})/\sqrt{2}+O(l/\lambda_R),\\
|3\rangle&=&(-|00\downarrow\rangle+|10\uparrow\rangle)/\sqrt{2}+O(l/\lambda_R).
\end{eqnarray*}
 Note that, although for a GaAs QD the level anticrossing is a quite small effect, the mixing of spin-up and spin-down states occurs in a sufficiently large region of magnetic fields (for a GaAs QD with $\hbar\omega_0=1.1\:$meV and $\lambda_R=8\:\mu$m, the mixing occurs essentially in the region of width $\approx1\:$Tesla)  and thus can be observed experimentally. Indeed, let us consider relaxation processes between the state $|1\rangle$ (spin-up) and the states $|2\rangle$, $|3\rangle$. Beyond the mixing region one of the latter states is spin-up, the other is spin-down. Spin-flip relaxation is much slower than orbital relaxation \cite{FTH}, therefore, relaxation to the ground state from the states $|2\rangle$ and $|3\rangle$ is very different. However, in the region of mixing of spin-up and spin-down states, the spin-flip relaxation strongly increases and becomes comparable with orbital relaxation. Note that these anticrossing features in semiconductor QDs are very similar to the ``hot spots'' in polyvalent metals \cite{ZFS}.

Moreover, it is interesting to note that spin relaxation due to the Rashba SO coupling differs from that due to the Dresselhaus SO coupling in this mixing region. As mentioned above, in the case of the Dresselhaus term there is no SO interaction between the states $|2\rangle$ and $|3\rangle$ [see inset in Fig.~\ref{fig:1}(a)], therefore, there is no spin mixing of these states. Thus, spin relaxation due to the Dresselhaus SO coupling does not undergo a considerable increase, in contrast to spin relaxation due to the Rashba SO coupling. Note that, in the general case, when the SO coupling includes both the Rashba and Dresselhaus terms, there is no interplay between the Dresselhaus and Rashba terms in the spin relaxation rate in perpendicular magnetic fields \cite{GKL} and the total rate is just the sum of two terms caused by these SO couplings. Therefore, we can study these two terms separately.

\section{Spin relaxation}
We consider next phonon-induced relaxation in a QD. The coupling between electrons and phonons  with mode $\mathbf{k}j$ ($\mathbf{k}$ is the phonon wave vector and $j$ is the branch index $j=L,\ T1,\ T2$ for one longitudinal and two transverse modes) is given by\cite{GKL}
\begin{equation}
\label{eq:U_ph}
U_{\mathbf{k}j}^{ph}(\mathbf{r})=\sum_j\frac{F(k_z)}{\sqrt{2\rho V k s_j/\hbar}}(eA_{\mathbf{k}j}-ik\Xi_{\mathbf{k}j})e^{i\mathbf{kr}}b^+_{\mathbf{k}j}+c.c.,
\end{equation}
where $\rho$ is the crystal mass density, $V$ is the volume of the QD,  $s_j$ is the sound velocity, 
$A_{\mathbf{k}j}=\xi_i\xi_ld^{\mathbf{k}j}_m\beta_{ilm}$, $\bm{\xi}=\mathbf{k}/k$,
$\mathbf{d}^{\mathbf{k}j}$ is the phonon polarization vector, $\Xi_{\mathbf{k}j}$ is the deformation potential, and $\beta_{ilm}$ is the piezotensor, which has
nonzero components only when all three indices $i,l,m$ are different: $\beta_{xyz}=\beta_{xzy}=\ldots=h_{14}/\varepsilon_S$
($\varepsilon_S$ is the static dielectric constant). 
For GaAs, $eh_{14}=1.2\times10^7\:$eV$/$cm, $\varepsilon_S=13.2$,  $\Xi_{\mathbf{k}j}=\delta_{j,L}\Xi_0$, and $\Xi_0=6.7\:$eV). In Eq.~(\ref{eq:U_ph}) we introduced the form-factor $F(k_z)$ which is determined by the spread of the electron wave function
 in the $z$-direction: $F(k_z)=\int dze^{ik_zz}|\psi_0(z)|^2$, where $\psi_0(z)$ is the ground state envelope wave
function of an electron along the $z$-direction. The form factor $F(k_z)$ equals unity for $|k_z|\ll d^{-1}$ and vanishes for
$|k_z|\gg d^{-1}$ (see Ref.~\onlinecite{GKL}).

 Let us find contributions to the relaxation rate of transitions between the levels $|1\rangle$ and $|2\rangle$ ($\Gamma_{21}$); $|2\rangle$ and $|3\rangle$ ($\Gamma_{32}$); $|1\rangle$ and $|3\rangle$ ($\Gamma_{31}$). In the framework of the Bloch -- Redfield theory, the phonon-induced relaxation rate ($1/T_1$) of a two-level system is a sum of transition probabilities between levels accompanied by absorption and emission of phonons \cite{Blum} and, for a QD, the decoherence time is $T_2=2T_1$ \cite{GKL}.
Therefore, using Fermi's golden rule and the expressions for the three lowest levels with the Dresselhaus SO coupling [Eqs.~(\ref{eq:State2D}) and (\ref{eq:State3D})], we get  the rates
\begin{eqnarray}\nonumber
\Gamma_{21}&=&\frac{l^4\omega_Z^3\left(N_{\omega_Z}+1/2\right)}{8\pi\hbar\rho\lambda^2_D}\left(\frac{\omega_1}{\omega_1+\omega_Z} -\frac{\omega_2}{\omega_2-\omega_Z}\right)^2\\
&&\times\sum_js_j^{-5}e^{-\omega_Z^2l^2/2s^2_j} I_j^{(3)}\left(\omega_Z\right),\\
\nonumber
\Gamma_{32}&=&\frac{w^5l^6\left(N_w+1/2\right)}{32\pi\hbar\rho\lambda_D^2}\left(\frac{\omega_1}{\omega_1+\omega_Z} -\frac{\omega_2}{\omega_2-\omega_Z}\right)^2\\
&&\times\sum_js_j^{-7}e^{-w^2l^2/2s^2_j} I_j^{(5)}\left(w\right),\\
\Gamma_{31}&=&\frac{\omega_1^3\left(N_{\omega_1}+1/2\right)}{8\pi\rho m^*\Omega}\sum_js_j^{-5}e^{-\omega_1^2l^2/2s^2_j} I_j^{(3)}\left(\omega_1\right),
\end{eqnarray}
where $w=\omega_1-\omega_Z$, $N_w=\left(e^{\hbar w/T}-1\right)^{-1}$, and
\begin{eqnarray}
\nonumber
I^{(m)}_j(\omega)&=&\int_0^{2\pi}d\varphi\int_0^{\pi/2}d\vartheta\sin^m\vartheta 
e^{\omega^2l^2\cos^2\vartheta/2s_j^2}\\
\label{eq:I}
&&\times F^2(\omega\cos\vartheta/s_j)\left[\left(eA_{\mathbf{k}j}\right)^2+\frac{\omega^2}{s_j^2}\delta_{j,L}\Xi_0^2\right].
\end{eqnarray}
In the case of parabolic confinement along the growth direction of a QD, $I^{(m)}_j(x)$ can be expressed in terms of error functions (see \appendixname~\ref{app:1}).

% Here, for brevity, we present only the part of $F_L(a)$ and $G_L(a)$ which are due to deformational acoustic (DA) phonons and which are dominant in the relaxation processes at high magnetic fields \cite{GKL,AFM}.

In the case of Rashba SO coupling alone, we have
\begin{eqnarray}
\nonumber
\Gamma_{21}&=&\frac{{w_-}^3\left(N_{w_-}+1/2\right)}{8\pi\rho m^*\Omega}\left(\sin\gamma/2+ \frac{(l/\lambda_R)\omega_2}{\omega_2+\omega_Z}\cos\gamma/2\right)^2\\
&&\times\sum_js_j^{-5}e^{-{w_-}^2l^2/2s^2_j} I_j^{(3)}\left(w_-\right),\\
\nonumber
\Gamma_{32}&=&\frac{\omega_R^5\hbar\left(N_{\omega_R}+1/2\right)}{32\pi\rho(m^*\Omega)^2}\left(\frac12 \sin\gamma+\frac{(l/\lambda_R)\omega_2}{\omega_2+\omega_Z}\cos\gamma\right)^2\\
&&\times\sum_js_j^{-7}e^{-\omega_R^2l^2/2s^2_j} I_j^{(5)}\left(\omega_R\right),\\
\nonumber
\Gamma_{31}&=&\frac{{w_+}^3\left(N_{w_+}+1/2\right)}{8\pi\rho m^*\Omega}\left(\cos\gamma/2- \sin\gamma/2\frac{(l/\lambda_R)\omega_2}{\omega_2+\omega_Z}\right)^2\\
&&\times\sum_js_j^{-5}e^{-{w_+}^2l^2/2s^2_j} I_j^{(3)}\left(w_+\right),
\end{eqnarray}
where $w_\pm=(\omega_1+\omega_Z\pm\omega_R)/2$ and $\omega_R$ is defined by Eq.~(\ref{eq:wR}).

\section{Analysis and discussion}
Figure~\ref{fig:1} shows these contributions to the relaxation rate due to the Dresselhaus and the Rashba SO couplings. As can be seen from this figure, the orbital relaxation rate (the dashed and dot-dashed curves) is independent of the SO coupling. The behavior of the spin relaxation rate $\Gamma_{32}$ (the dotted curves) is qualitatively the same for both the Dresselhaus and the Rashba SO couplings. Solid curves correspond to the 
the spin relaxation rate $\Gamma^{\downarrow\uparrow}$ between the Zeeman-split orbital ground state levels ($\Gamma^{\downarrow\uparrow}=\Gamma_{21}$ for the Dresselhaus SO coupling and in the case of the Rashba SO coupling $\Gamma^{\downarrow\uparrow}=\Gamma_{21}$ on the left side of the cusp and $\Gamma^{\downarrow\uparrow}=\Gamma_{32}$ on the right of the cusp). Significantly,  $\Gamma^{\downarrow\uparrow}$, in the case of the Rashba SO coupling, possesses a cusplike structure at the anticrossing point \cite{footnote3}, whereas, in the case of Dresselhaus SO coupling, $\Gamma^{\downarrow\uparrow}$ is a monotonic function of $B$ \cite{footnote5,footnote6}. 

It should be noted that at $B>1\:$T the relaxation due to deformational acoustic (DA) coupling is much faster than that due to piezoelectric (PE) coupling, except in the case of orbital relaxation at high magnetic fields, when relaxation induced by PE-phonons is of the same order as that due to DA-phonons. 
Since $d\ll l_0$ and $q\approx l_0^{-1}$, the factor  $F(q_z)\approx1$  in the electron-phonon interaction operator (see Ref.~\onlinecite{GKL}) and the relaxation is practically independent of $d$ aside from the orbital relaxation at low magnetic fields: $\Gamma_{31}(B=0)\approx \omega_0^4(N_{\omega_0}+1/2)\Xi_0^2e^{-\omega_0^2d^2/2s_1^2} / 6\pi\rho m^*s_1^7$ (the spin relaxation rates are zero at $B=0$). 
The orbital relaxation rate has a maximum when the phonon wave length is comparable to the lateral size $l$ of a QD ($ql\approx3$). At high magnetic fields, the orbital relaxation rate decreases with $B$ [as $(\omega_0/\omega_c)^6$ for DA coupling and as $(\omega_0/\omega_c)^4$ for PE coupling], since $\omega_1\to\omega_0^2/\omega_c$ at high $B$. The rate $\Gamma_{32}\propto\omega_Z^2(\omega_1-\omega_Z)^5$ at low magnetic fields, is zero at the anticrossing, and $\Gamma_{32}\propto\omega_Z^4$ at high magnetic fields. The spin relaxation rate between the Zeeman-split levels $\Gamma^{\downarrow\uparrow}\propto\omega_Z^k$ (at low magnetic fields $k=7$ for DA coupling and $k=5$ for PE coupling. At high $B$, $k=3$ for DA coupling and $k=1$ for PE coupling). In the anticrossing region, the spin relaxation rate due to the Dresselhaus SO coupling is a monotonic function of $B$: $\Gamma^{\downarrow\uparrow}=\Gamma_{21}\propto\omega_Z^3$, but that due to the Rashba SO coupling has a strong increase at the anticrossing point and near this point $\Gamma^{\downarrow\uparrow}\propto\omega_Z^3/[(1-\omega_Z/\omega_1)^2+4(l/\lambda_R)^2]$. 
Therefore, there is both a qualitative difference (in the magnetic-field dependence)
and quantitative difference (at $4.8\:$T the Rashba SO coupling gives $\Gamma^{\downarrow\uparrow}\approx10^4\:$s$^{-1}$ but the Dresselhaus SO coupling gives $\Gamma^{\downarrow\uparrow}\approx70\:$s$^{-1}$) in  the behavior of the spin relaxation rate $\Gamma^{\downarrow\uparrow}$ due to the Dresselhaus and Rashba SO coupling. This can serve as a means of extracting information on the different contributions to the total SO coupling strength \cite{footnote4}.

Note that, with a decrease in the lateral size  $l_0$ of a QD, the cusp and the maximum in the orbital relaxation rate are shifted to high magnetic fields. For a larger SO coupling (smaller SO length), the spin relaxation rates have higher values, because $\Gamma^{\downarrow\uparrow}\propto\lambda_{so}^{-2}$, and the cusp shape is smoother. The temperature dependence of the relaxation rates is only important for transitions between the levels with a separation comparable to the temperature: the rates decrease with temperature for the orbital relaxation at high magnetic fields (when the level spacing $\sim\hbar\omega_0^2/\omega_c$), for the spin relaxation between Zeeman-split levels (when the Zeeman energy $\sim T$) at low magnetic fields, and for $\Gamma_{32}$ at the anticrossing (when the level spacing $|\omega_1-\omega_Z|\sim T$).

\section{Conclusions}

We have shown that at an accidental degeneracy point the Rashba SO coupling leads to an anticrossing. The mixing of the spin-up and spin-down states at the anticrossing enhances the spin relaxation rate due to the Rashba SO coupling relative to the spin relaxation rate due to the Dresselhaus SO coupling.

\begin{acknowledgments} The authors thank V.N.~Golovach, W.A.~Coish, and J.~Lehmann for useful discussions.
The authors acknowledge support from the Swiss NSF, NCCR Basel, EU RTN ``Spintronics'', U.S. DARPA, ARO, and ONR.
\end{acknowledgments}

\appendix
\section{Parabolic confinement along the $z$-direction}
\label{app:1}
In the case of parabolic confinement along the growth direction of a QD, $F(k_z)=\exp(-d^2k_z^2/4)$, where $d$ is the width of the quantum well, and integrals in Eq.~(\ref{eq:I}) can be expressed in terms of the imaginary error functions $\mathrm{erfi}(x)$. After some algebra we get
\begin{widetext}
\begin{eqnarray*}
\label{eq:ILA}
I_{L}^{(3)}(a/\tau_L)&=&\left(\frac{eh_{14}}{\varepsilon_S}\right)^2\frac{9}{4a^4}\left[e^{a^2}\left(1+\frac{5}{a^2}+\frac{105}{4a^4}\right)-\sqrt{\pi}\mathrm{erfi}(a)
\left(a+\frac{9}{2a}+\frac{45}{4a^3}+\frac{105}{8a^5}\right)\right]\\
&&+\Xi_0^2[-2e^{a^2}+\sqrt{\pi}\mathrm{erfi}(a)(1/a+2a)]/(l^2-d^2),\\
I_{T1}^{(3)}(a/\tau_{T1})&=&\left(\frac{eh_{14}}{\varepsilon_S}\right)^2\frac{1}{a^4}\left[e^{a^2}\left(1+\frac{15}{2a^2}\right)-\sqrt{\pi}\mathrm{erfi}(a)
\left(a+\frac{3}{a}+\frac{15}{4a^3}\right)\right],\\
\nonumber
I_{T2}^{(3)}(a/\tau_{T2})&=&\left(\frac{eh_{14}}{\varepsilon_S}\right)^2\frac{1}{4a^4}\left[-e^{a^2}\left(2a^2+9+\frac{45}{2a^2}+\frac{945}{4a^4}\right)+\sqrt{\pi}\mathrm{erfi}(a)
\left(2a^3+8a+\frac{33}{a}+\frac{90}{a^3}+\frac{945}{8a^5}\right)\right],\\
\label{eq:JLA}
I_{L}^{(5)}(a/\tau_{L})&=&\left(\frac{eh_{14}}{\varepsilon_S}\right)^2\frac{9}{4a^4}\left[e^{a^2}\left(1+\frac{13}{2a^2}+\frac{105}{4a^4}+\frac{945}{8a^6}\right)-\sqrt{\pi}\mathrm{erfi}(a)
\left(a+\frac{6}{a}+\frac{45}{2a^3}+\frac{105}{2a^5}+\frac{945}{16a^7}\right)\right]\\
&&+\Xi_0^2[-(3 + 2a^2) e^{a^2}+ \sqrt{\pi}\mathrm{erfi}(a)
(3/2a + 2a + 2a^3)] /a^2(l^2-d^2),\\
I_{T1}^{(5)}(a/\tau_{T1})&=&\left(\frac{eh_{14}}{\varepsilon_S}\right)^2\frac{1}{a^4}\left[e^{a^2}\left(1+\frac{5}{a^2}+\frac{105}{4a^4}\right)-\sqrt{\pi}\mathrm{erfi}(a)
\left(a+\frac{9}{2a}+\frac{45}{4a^3}+\frac{105}{8a^5}\right)\right],\\
\nonumber
I_{T2}^{(5)}(a/\tau_{T2})&=&\left(\frac{eh_{14}}{\varepsilon_S}\right)^2\frac{1}{4a^4}\left[-e^{a^2}\left(2a^2+10+\frac{51}{a^2}+\frac{315}{2a^4}+\frac{8505}{8a^6}\right)\right.\\
\label{eq:JTA2}
&&\left.+\sqrt{\pi}\mathrm{erfi}(a)
\left(2a^3+9a+\frac{45}{a}+\frac{345}{2a^3}+\frac{3465}{8a^5}+\frac{8505}{16a^7}\right)\right],
\end{eqnarray*}
where $\tau_j^2=(l^2-d^2)/2s_j^2$.
\end{widetext}


\begin{thebibliography}{99}
\bibitem{footnote} We neglect the cubic Dresselhaus term which is important for 2D quantum wells \cite{Miller}, 
but for a few-electron QD  with the lateral size greater than the height ($l_0\gg d$) the 
cubic term is smaller than the linear one by a factor of $(d/l_0)^2$.
\bibitem{footnote3} In the case of the Rashba SO coupling, our treatment is valid only if $\Delta>\hbar/T_1$ (for a GaAs QD with $\hbar\omega_0=1.1\:$meV and for the Rashba SO coupling $\lambda_R=8\:\mu$m, $\Delta\approx3\hbar/T_1$). Otherwise the secular approximation  \cite{Blum} breaks down.
\bibitem{footnote5}Due to the time-reversal symmetry of the Hamiltonian the behavior of the 
spin relaxation rate  $\Gamma^{\downarrow\uparrow}$ does not change with the sign 
change of the magnetic field. But the inversion of the sign of the $g$-factor leads to opposite 
effects: the Dresselhaus SO coupling now gives rise to anticrossing and to the cusplike feature 
of $\Gamma^{\downarrow\uparrow}$, while the Rashba term does not.
\bibitem{footnote6}
In the case of anisotropic 2D QDs characterized by $\omega_x\ne\omega_y$, there is also mixing of the states due to Dresselhaus coupling, which depends on the strength of anisotropy. For weak anisotropy $[(\omega_x-\omega_y)/\omega_c\ll1]$, this leads to a narrow cusp in the spin relaxation which can be neglected with respect to that due to Rashba coupling. Note that anisotropy effects in QDs in strong magnetic fields are studied in Ref.~\onlinecite{FAT}
\bibitem{footnote4} One of the possible experimental separations of the SO constants $\alpha$ and $\beta$ 
could be the following: From the width of the cusp of $\Gamma^{\downarrow\uparrow}$ the Rashba 
constant $\alpha$ can be found. Then taking into account that the relaxation rate is just a sum 
of the two SO contributions,  we can find the Dreesselhaus SO coupling $\beta$ in the region of 
magnetic field away from the cusp.
\end{thebibliography}
\end{document}